\documentclass[aps,prb,twocolumn,showpacs,superscriptaddress,longbibliography]{revtex4-2}
\usepackage{amsfonts}
\usepackage{amsmath}
\usepackage{graphicx}
\usepackage{bm}
\usepackage{amssymb}
\usepackage{dcolumn}
\usepackage{color}
\usepackage{multirow}
\usepackage{booktabs}
\usepackage[colorlinks,
linkcolor=blue,
anchorcolor=blue,
citecolor=blue,
urlcolor=blue,
]{hyperref}

\begin{document}

    \title{Room temperature ferromagnetic semiconductors through metal-semiconductor transition in monolayer MnSe$_2$}

    \author{Jia-Wen Li}
    \affiliation{Kavli Institute for Theoretical Sciences, University of Chinese Academy of Sciences, Beijng 100049, China}

    \author{Gang Su}
    \email{gsu@ucas.ac.cn}
    \affiliation{Kavli Institute for Theoretical Sciences, University of Chinese Academy of Sciences, Beijng 100049, China}
    \affiliation{CAS Center for Excellence in Topological Quantum Computation, University of Chinese Academy of Sciences, Beijng 100190, China}
    \affiliation{Physical Science Laboratory, Huairou National Comprehensive Science Center, Beijing 101400, China}
    \affiliation{School of Physical Sciences, University of Chinese Academy of Sciences, Beijng 100049, China}

    \author{Bo Gu}
    \email{gubo@ucas.ac.cn}
    \affiliation{Kavli Institute for Theoretical Sciences, University of Chinese Academy of Sciences, Beijng 100049, China}
    \affiliation{CAS Center for Excellence in Topological Quantum Computation, University of Chinese Academy of Sciences, Beijng 100190, China}
    \affiliation{Physical Science Laboratory, Huairou National Comprehensive Science Center, Beijing 101400, China}

    \begin{abstract}
        To realize room temperature ferromagnetic semiconductors is still a challenge in spintronics.
        Recent experiments have obtained two-dimensional (2D) room temperature ferromagnetic metals, such as monolayer MnSe$_2$.
        In this paper, we proposed a way to obtain room temperature ferromagnetic semiconductors through metal-semiconductor transition.
        By the density functional theory calculations, a room temperature ferromagnetic semiconductor is obtained in monolayer MnSe$_2$ with a few percent tensile strains, where a metal-semiconductor transition occurs with 2.2\% tensile stain. 
        The tensile stains raise the energy of d orbitals of Mn atoms and p orbitals of Se atoms near the Fermi level, making the Fermi level sets in the energy gap of bonding and antibonding states of these p and d orbitals, and opening a small band gap.   
        The room temperature ferromagnetic semiconductors are also obtained in the heterostructures MnSe$_2$/X (X = Al$_2$Se$_3$, GaSe, SiH, and GaP), where metal-semiconductor transition happens due to the tensile strains by interface of heterostructures.  
        In addition, a large magneto-optical Kerr effect (MOKE) is obtained in monolayer MnSe$_2$ with tensile strain and MnSe$_2$-based heterostructures.
        Our theoretical results pave a way to obtain room temperature magnetic semiconductors from experimentally obtained 2D room temperature ferromagnetic metals through metal-semiconductor transitions.
    \end{abstract}
    \pacs{}
    \maketitle

    %%%%%%% Main text %%%%%%%%%%%%%%%%%%%%%

    \section{Introduction}

    In spintronics, it is still a challenge in experiments to realize room temperature ferromagnetic semiconductors.
    In 2017, the successful synthesis of two-dimensional (2D) van der Waals ferromagnetic semiconductors CrI$_3$ \cite{Huang2017} and Cr$_2$Ge$_2$Te$_6$ \cite{Gong2017} in experiments has attracted extensive attention to 2D ferromagnetic semiconductors.
    According to Mermin-Wagner theorem \cite{Mermin1966}, the magnetic anisotropy is essential to produce long-range magnetic order in 2D systems. Recently, more 2D ferromagnetic semiconductors have been obtained in experiments, such as Cr$_2$S$_3$ \cite{Chu2019}, CrCl$_3$ \cite{Cai2019}, CrBr$_3$ \cite{Zhang2019}, CrSiTe$_3$ \cite{Achinuq2021}, CrSBr \cite{Lee2021}, where the Curie temperatures $T_C$ are far below room temperature.
    On the other hand, the 2D van der Waals
    ferromagnetic metals with high $T_C$ have been obtained in
    recent experiments.
    For example, $T_C$ = 140 K in CrTe \cite{Sun2022}, 300 K in CrTe$_2$ \cite{Meng2021}, 344 K in Cr$_3$Te$_6$ \cite{Chua2021}, 160 K in Cr$_3$Te$_4$ \cite{Li2022b}, 280 K in CrSe \cite{Zhang2019a}, 300 K in Fe$_3$GeTe$_2$ \cite{Deng2018,Fei2018}, 270 K in Fe$_4$GeTe$_2$ \cite{Seo2020}, 229 K in Fe$_5$GeTe$_2$ \cite{May2019}, 380 K in Fe$_3$GaTe$_2$ \cite{Zhang2022}, 300 K in MnSe$_2$ \cite{O’hara2018}, etc.

    Recent studies have shown that the physical properties of 2D materials are sensitive to external regulations, such as electric filed \cite{Deng2018,Jiang2018,You2021,Zhao2021,Shi2019}, doping \cite{Hao2019,Wang2018,Feng2017,Fang2022,Zhao2022a}, surface functionalization \cite{GonzalezHerrero2016,Datta2017}, intercalation \cite{Pathirage2023,Zhou2021a}, heterostructure \cite{Chen2019,Zhang2021,Kumar2016,Chen2023,Wu2023,Dong2020,Chen2019a,Li2023,Ren2016a,Caglayan2022,Yuan2022,Yuan2023,Hong2022,Yang2023,Zhu2023,Lu2023}, strain \cite{Huang2019,Li2020,Xu2018,Wu2019,Zhang2021,Wei2023}, etc. Among them, strain engineering is an effective technique to change the lattice and electronic structures and thus to control various properties of 2D materials.
    In contrast to bulk materials, 2D materials have stronger deformation capacity and thus can withstand greater elastic strain without fracture, which shows great advantages in strain engineering.
    For example, monolayer MoS$_2$ can sustain strains as large as 11\% \cite{Bertolazzi2011}, monolayer FeSe can sustain strains up to 6\% \cite{Peng2014,Zhang2016}, and single-layer graphene can even withstand 25\% elastic strain \cite{Cocco2010}.
    It has been observed that in few-layer black phosphorus that biaxial tensile strain can increases the band gap, while biaxial compressive strain can decrease the band gap \cite{Huang2019}.
    Under uniaxial strain up to 1.7\%, a band gap reduction of 0.3 eV was observed in monolayer MoS$_2$ \cite{Li2020}.

	In addition, heterostructures can have a significant impact on the magnetism of magnetic materials \cite{Dong2020,Chen2019a,Li2023,Hong2022,Yang2023,Zhu2023}.
	Heterostructures also have a great impact on the band structure. 
	Especially, due to the effect of interlayer interaction and strain, metal-semiconductor transition happens in some
	semimetal/semiconductor heterostructures, and a band gap $E_g$ can be opened, such as $E_g$ = 0.44 eV in silicene/GaP \cite{Ren2016a}, 0.22 eV in silicene/Ga$_2$SeS \cite{Caglayan2022}, 0.67 eV in CSe/BP \cite{Idrees2023}, 13 meV in graphene/MoSi$_2$N$_4$ \cite{Yuan2022} and 17 meV in Ge/SMoSe \cite{Yuan2023}.
	
    As a room temperature ferromagnetic material, MnSe$_2$ has attracted a lot of attentions due to its interesting properties \cite{Hu2021,Chen2023,Wu2023,Noesges2020,Xiao2023,Eren2019,Xie2021,Li2022,Kan2014,He2022}.
    The structure of MnSe$_2$ is shown in Fig. \ref{fig_MnSe2}(a) with space group of P$\overline{3}$m1 (164).
    The band structure from density functional theory (DFT) with Perdew-Burke-Ernzerhof (PBE) pseudopotential shows semimetal properties \cite{Eren2019,Xie2021,Li2022,Kan2014}.
    Because PBE pseudopotential always underestimated the band gap, Heyd-Scuseria-Ernzerhof (HSE) hybrid functional approach is believed to give a better description of the band structures \cite{Heyd2003}.
    By the DFT with HSE, MnSe$_2$ was calculated to be a semimetal \cite{Xie2021,Hong2022} or semiconductor with small band gap of 0.01 eV \cite{Kan2014}.
    Experimental results suggest that MnSe$_2$ at Bi$_2$Se$_3$ substrate has a small band gap \cite{Noesges2020}.
    On the other hand, the electronic properties of monolayer MnSe$_2$ are still unclear.
    It was reported that tensile strain could enhance the ferromagnetic properties of monolayer MnSe$_2$ and weaken the metallicity \cite{Xie2021,Kan2014}.
    It has been discussed that $T_C$ can be enhanced by tensile strain in some 2D ferromagnetic semiconductors \cite{Dong2019,Dong2020,Zhang2021b,O’Neill2022}
    %Due to the semi-metallic properties, the electronic properties of MnSe$_2$ under tensile strain is important.
    The evolution of band structures of monolayer MnSe$_2$ was calculated by DFT with PBE, giving no band gap with tensile strain up to 8\% \cite{Kan2014}.
    Is it possible to open a band gap in these room temperature ferromagnetic metals by some methods, and obtain room temperature ferromagnetic semiconductors?
    %Is it practicable to obtain room temperature ferromagnetic semiconductors by adding tensile strain under MnSe$_2$?

    In this paper, we propose a way to obtain room temperature ferromagnetic semiconductor by metal-semiconductor transition in monolayer MnSe$_2$ and MnSe$_2$-based heterostructures. 
    The DFT calculation with HSE hybrid functional shows that monolayer MnSe$_2$ is room temperature ferromagnetic semimetal.
    A 2.2\% tensile strain could open a small band gap in monolayer MnSe$_2$.
    An in-plane to out-of-plane magnetic anisotropy transition happens with 1.7\% tensile strain, and $T_C$ increases with tensile strain.
    In addition, the heterostructures MnSe$_2$/X with X= GaP \cite{Li2022a,Parmar2022,Wei2022}, GaSe \cite{Li2023a}, SiH \cite{Wani2022}, and Al$_2$Se$_3$ \cite{Li2023} were studied. 
    The room temperature ferromagnetic semiconductors are also obtained in these heterostructures, where metal-semiconductor transition occurs due to tensile stain by interfaces.
    Large magneto-optical Kerr effect is found in monolayer MnSe$_2$ with a percent tensile strain and MnSe$_2$-based heterostructures.
    Our theoretical results propose a way to obtain room temperature ferromagnetic semiconductors by metal-semiconductor transition in monolayer MnSe$_2$ by applying tensile strain or building heterostructures.

    \section{Method}
    All calculations were based on the DFT as implemented in the Vienna ab initio simulation package (VASP) \cite{Kresse1996}.
    The exchange-correlation potential is described by the PBE form with the generalized gradient approximation (GGA) \cite{Perdew1996}.
    The electron-ion potential is described by the projector-augmented wave (PAW) method \cite{Bloechl1994}.
    We carried out the calculation of PBE + U with U = 4 eV for 3d electrons in Mn \cite{Eren2019,Xie2021,Li2022,Kan2014}.
    The band structures were calculated in HSE06 hybrid functional \cite{Heyd2003}.
    The plane-wave cutoff energy is set to be 650 eV.
    The 9$\times$9$\times$1 $\Gamma$ center K-point was used for the Brillouin zone (BZ) sampling.
    To obtain accurate results of magnetic anisotropy energy  (MAE), K-points were chosen as $\Gamma$-centered 25$\times$25$\times$1.
    The density of states (DOS) were obtained from HSE and  K-points of $\Gamma$-centered 18$\times$18$\times$1.
    The structures of all materials were fully relaxed, where the convergence precision of energy and force was $10^{-6}$ eV and $10^{-2}$ eV/Å, respectively.
    The van der Waals effect is included with DFT-D3 method \cite{Grimme2010}.
    The Wannier90 code was used to construct a tight-binding Hamiltonian \cite{Mostofi2008,Mostofi2014}.
    The WannierTools code was used to obtain a tight-binding band structure from tight-binding Hamiltonian \cite{Wu2018}.
    The Heisenberg-type Monte Carlo simulation was performed on a 50$\times$50$\times$1 lattice with 2500 magnetic points for MnSe$_2$.
    1$\times$10$^5$ steps were carried for each temperature, and the last one-thirds steps were used to calculate the temperature-dependent physical quantities.

    \section{RESULTS AND DISCUSSION}
	
    \subsection{Monolayer MnSe$_2$}
    
\begin{figure}[tphb]
		\centering
		\includegraphics[scale=0.4,angle=0]{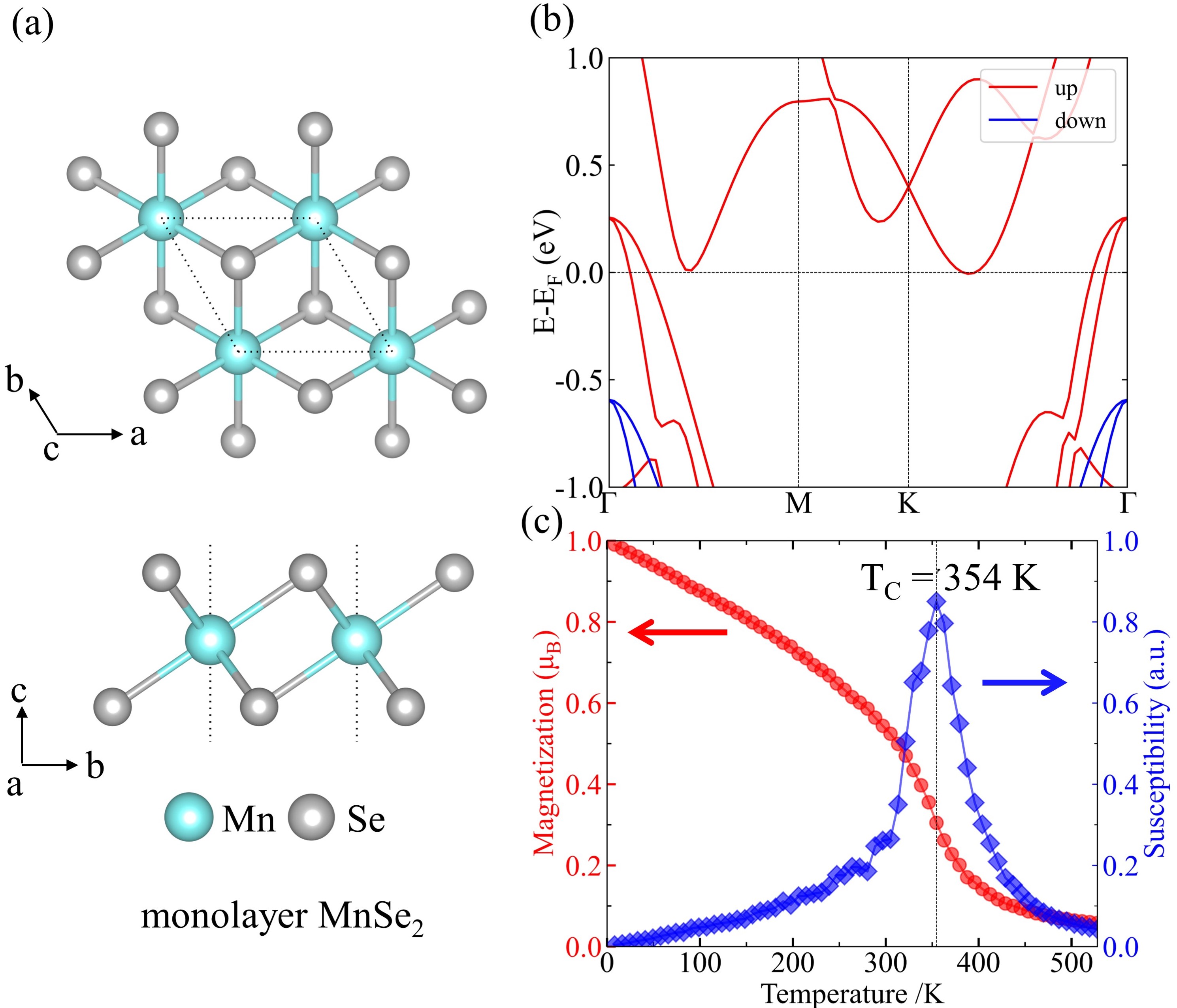}\\
		\caption{
			(a) Top and side views of crystal structure of monolayer MnSe$_2$. 
			(b) Spin polarized band structure of MnSe$_2$, obtained by the DFT calculation with HSE hybrid functional. 
			(c) Magnetization and susceptibility of MnSe$_2$ as a function of temperature, obtained by the Monte Carlo simulation based on a 2D Heisenberg model. 
			The Curie temperature of monolayer MnSe$_2$ is calculated as $T_C$ = 354 K, being consistent with the experiment \cite{O’hara2018}.
		}\label{fig_MnSe2}
	\end{figure}

    The crystal structure of monolayer MnSe$_2$ is shown in Fig. \ref{fig_MnSe2}(a), which shows a triangle lattice.
    The calculated in-plane lattice constant is $a_0$ =  3.658 \AA, in agreement with previous calculations \cite{Eren2019,Xie2021,Li2022,Kan2014}.
    The band structure of monolayer MnSe$_2$ with HSE hybrid functional is shown in Fig. \ref{fig_MnSe2}(b).
    The lowest band above Fermi level and the highest band below Fermi level slightly overlap, showing the semimetal behavior.
    The bands near Fermi level contain only one component of spins, showing the half-metal behavior.
    In addition, the Monte Carlo results of magnetization and susceptibility as a function of temperature for monolayer MnSe$_2$ is shown in Fig. \ref{fig_MnSe2}(c), giving a high $T_C$ = 354 K.
	
	\subsection{Monolayer MnSe$_2$ with strain}
	
	\begin{figure}[phbpt]
		\centering
		\includegraphics[scale=0.75,angle=0]{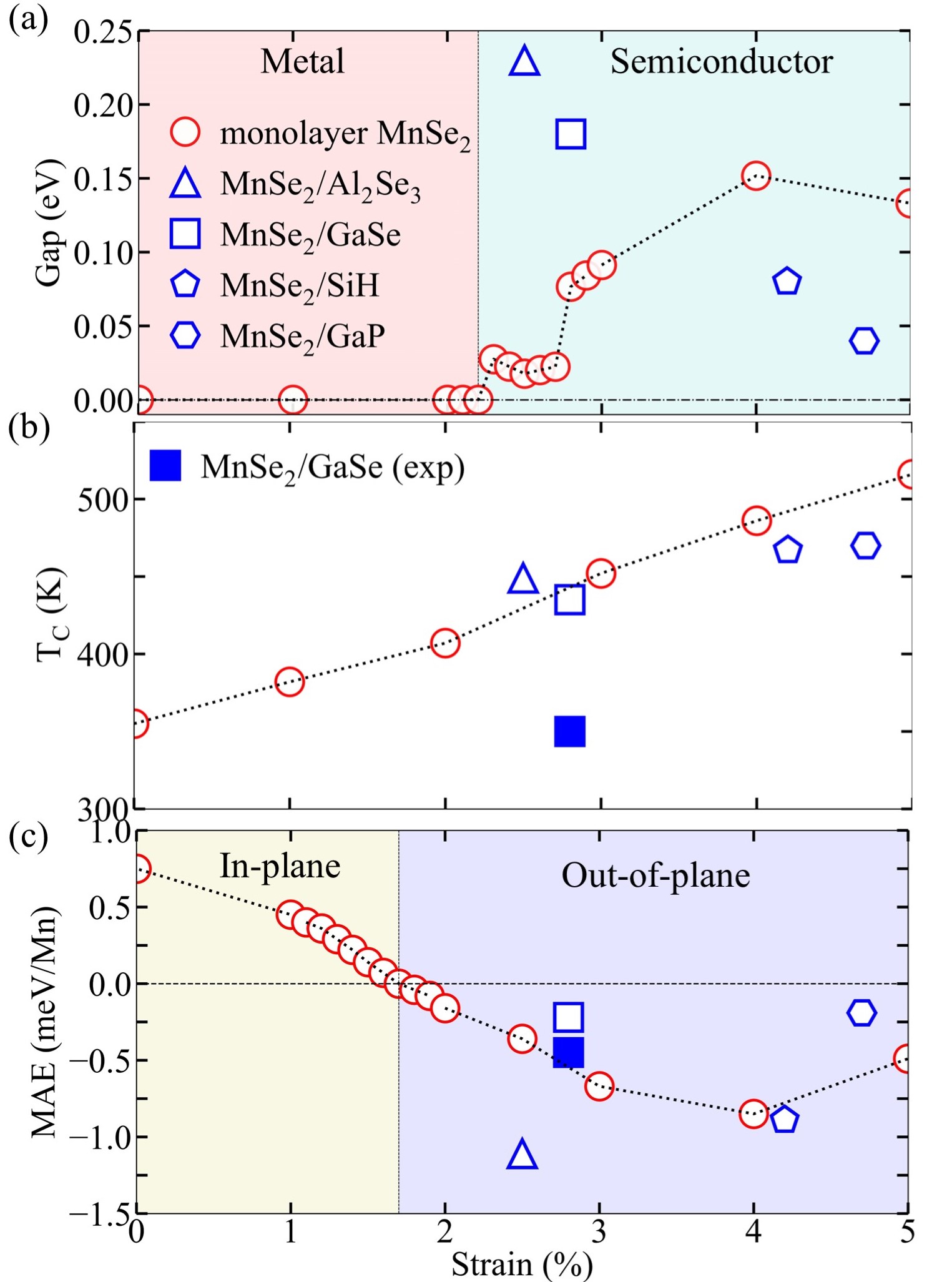}\\
		\caption{
			For monolayer MnSe$_2$, tensile strain dependence of (a) band gap, (b) Curie temperature $T_C$, and (c) magnetic anisotropy energy (MAE), obtained by the DFT and Monte Carlo calculations. 
			Numerical results of four MnSe$_2$-based heterostructures are also included, where strain comes from the interface of heterostructures.
			Experimental $T_C$ and MAE of heterostructure MnSe$_2$/GaSe \cite{O’hara2018} are included for comparison. 
		}\label{fig_Strain}
	\end{figure}
    To study the effect of strain on electronic properties, we applied biaxial strain. 
    The strain ratio is defined as $\epsilon=(a-a_0)/{a_0}$, where $a$ and $a_0$ are the in-plane lattice constants with and without strain, respectively.
    The variation of band gap of monolayer MnSe$_2$ with tensile strain calculated by DFT with HSE is shown in Fig. \ref{fig_Strain} (a).
    A metal-semiconductor transition happens with a tensile strain of 2.2\%.
    Fig. \ref{fig_Strain} (b) shows $T_C$ of monolayer MnSe$_2$ as a function with tensile strain. 
    $T_C$ increase with tensile strain, in agreement with previous report \cite{Dong2019,Dong2020,Zhang2021b,O’Neill2022}.
    The MAE is defined as $(E_{\perp}-E_{\parallel})/N_{Mn}$, where E$_{\perp}$ and E$_{\parallel}$ are energies of MnSe$_2$ with out-of-plane and in-plane magnetic polarization, respectively. 
    $N_{Mn}$ is the number of Mn atoms in a unit cell. For monolayer MnSe$_2$ without strain, as shown in Fig. \ref{fig_Strain}(c), the in-plane MAE of 0.75 meV/Mn is obtained, in agreement with the previous report \cite{Xie2021}.
    By applying tensile strain, an in-plane to out-plane MAE transition is obtained at 1.7\% tensile strain.
	Thus, monolayer MnSe$_2$ will become a room temperature ferromagnetic semiconductor with out-of-plane MAE by applying tensile strains above 2.2\%.
	
	\begin{figure}[phbpt]
		\centering
		\includegraphics[scale=0.4,angle=0]{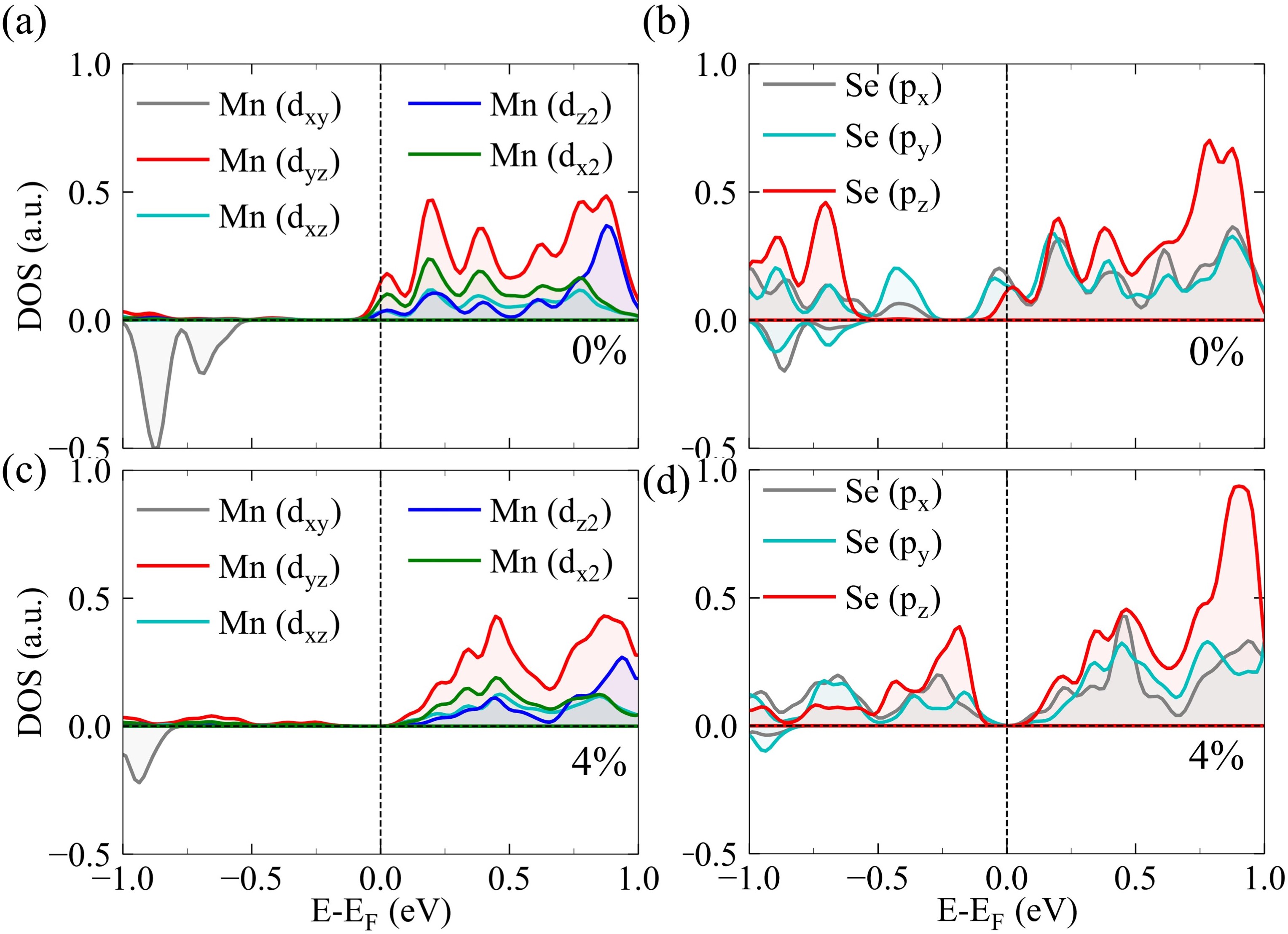}\\
		\caption{
			Metal-semiconductor transition of monolayer MnSe$_2$ due to strain. 
			DFT results of partial density of state (PDOS) of (a) d orbitals of Mn and (b) p orbitals of Se without stain. PDOS of (c) d orbitals of Mn and (d) p orbitals of Se with 4\% tensile stain. 
		}\label{fig_DOS}
	\end{figure}
	In order to analyze the metal-semiconductor transition of monolayer MnSe$_2$ under strain, the partial density of states (PDOS) were calculated, as shown in Fig. \ref{fig_DOS}.
	PDOS of d orbitals of Mn and p orbitals of Se without strain is shown in Figs. \ref{fig_DOS}(a) and \ref{fig_DOS}(b), respectively.
	PDOS of Mn and Se with 4\% tensile strain is shown in Figs. \ref{fig_DOS}(c) and \ref{fig_DOS}(d), respectively, with a small gap.
	The tensile stains raise the energy of d orbitals of Mn atoms and p orbitals of Se atoms near the Fermi level, making the Fermi level lie in the energy gap of bonding and antibonding states of these p and d orbitals, and opening a small band gap.

    \subsection{Heterostructures MnSe$_2$/X (X = GaP, GaSe, SiH, and Al$_2$Se$_3$)}
    
    Considering the mismatch from substrate is an efficient way to provide strain for 2D material in experiments \cite{Qi2023}, we constructed MnSe$_2$-based heterostructures with 2D semiconductors to provide strain.
    GaP \cite{Li2022a,Parmar2022}, GaSe \cite{Li2023a}, SiH \cite{Wani2022}, and Al$_2$Se$_3$ \cite{Li2023} are nonmagnetic semiconductors.
    The calculated results with PBE at monolayer limit give band gaps of 1.22, 1.80, 2.18 and 1.69 eV, respectively, and lattice constants of 3.916, 3.814, 3.888 and 3.788 \AA, respectively.
    According to the lattice mismatch $\delta = 2\times(a_1 -a_2)/(a_1+a_2) \times 100\%$, the lattice mismatch of GaP, GaSe, SiH, and Al$_2$Se$_3$ with MnSe$_2$ are 6.8\%, 4.2\%, 6.1\%, and 3.5\%, respectively.
    The detailed results are given in Supplemental Material \cite{SM}.

	\begin{figure*}[phbpt]
		\centering
		\includegraphics[scale=0.5,angle=0]{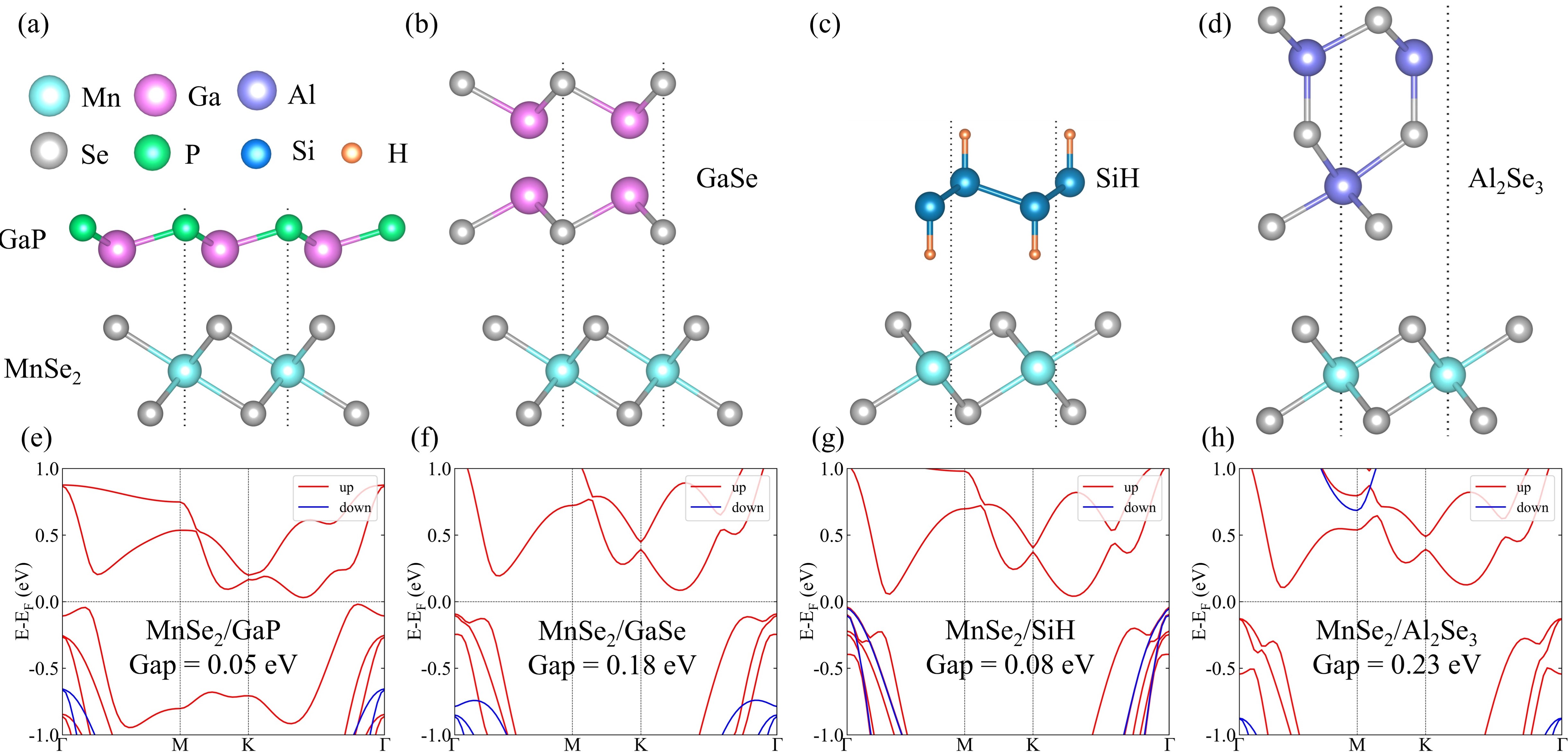}\\
		\caption{
			Structure and spin polarized band structure of MnSe$_2$-based heterostructures, (a, e) MnSe$_2$/GaP, (b, f) MnSe$_2$/GaSe, (c, g) MnSe$_2$/SiH, and (d, h) MnSe$_2$/Al$_2$Se$_3$. 
			The band structures are obtained by the DFT calculation with HSE hybrid functional.
		}\label{fig_bd}
	\end{figure*}
	
    We consider different stacking models of heterostructures, and the detailed data are given in Supplemental Material \cite{SM}.
    As shown in Fig. \ref{fig_bd}, all heterostructures are ferromagnetic semiconductors.
    The optimised lattice constants of MnSe$_2$/X with X = GaP, GaSe, SiH, and Al$_2$Se$_3$ are about 3.83, 3.76, 3.81, and 3.75 \AA, respectively, with an effective tensile strain of 4.7\%, 2.8\%, 4.2\%, and 2.5\%, respectively.
    The binding energy $E_b$ for MnSe$_2$/X is defined as $E_b=(E_{MnSe_2/X} - E_{MnSe_2} - E_{X})/N_{tot}$, where $E_{MnSe_2/X}$, $E_{MnSe_2}$, and $E_{X}$ represent the total energies of heterostructure MnSe$_2$/X, monolayer MnSe$_2$, and monolayer X, respectively, and $N_{tot}$ is the total atom number in a unitcell.
    The calculated results are -0.24, -0.17, -0.10, and -0.20 eV/atom for X = GaP, GaSe, SiH, and Al$_2$Se$_3$, respectively, indicating their stability.
    $T_C$ of heterostructures were obtained through DFT calculations and Monte Carlo simulation.
    As shown in Fig. 2, the calculation results predict that MnSe$_2$/X with X = GaP, GaSe, SiH, and Al$_2$Se$_3$ are room temperature magnetic semiconductors with out-of-plane MAE.
    It is noted that for the heterostructure MnSe$_2$/GaSe, Tc above room temperature and out-of-plane magnetization were observed in the experiment \cite{O’hara2018}. 
    Thus, our calculation results in Fig. \ref{fig_Strain} are consistent with the experiment of MnSe$_2$/GaSe \cite{O’hara2018}.
    The detailed calculation results are given in Supplemental Material \cite{SM}.
    
    Heterostructures not only provide effective strain, but also induce interlayer interaction.
    The calculated results of the most stable stacks of heterostructures is shown in Fig. \ref{fig_Strain}.
    The calculated results of heterostructures are different with those of monolayer MnSe$_2$ with the same tensile strain.
    For example, heterostructure MnSe$_2$/GaSe has a band gap of 0.18 eV, while monolayer MnSe$_2$ with same lattice constants has a band gap of 0.08 eV.
    Heterostructure MnSe$_2$/Al$_2$Se$_3$ give a MAE of -1.11 meV/Mn, while monolayer MnSe$_2$ with same lattice constants give a MAE of -0.36 meV/Mn.
    %In addition, MnSe$_2$ with 4.2\% tensile strain has no PBE band gap, while MnSe$_2$/GaP with the same lattice constants has a PBE band gap of 0.09 eV.
    Therefore, the interlayer interaction may play an important role in properties of heterostructures.

	\subsection{MOKE in monolayer MnSe$_2$ and heterostructures MnSe$_2$/X}

	\begin{figure}[phbpt]
		\centering
		\includegraphics[scale=0.36,angle=0]{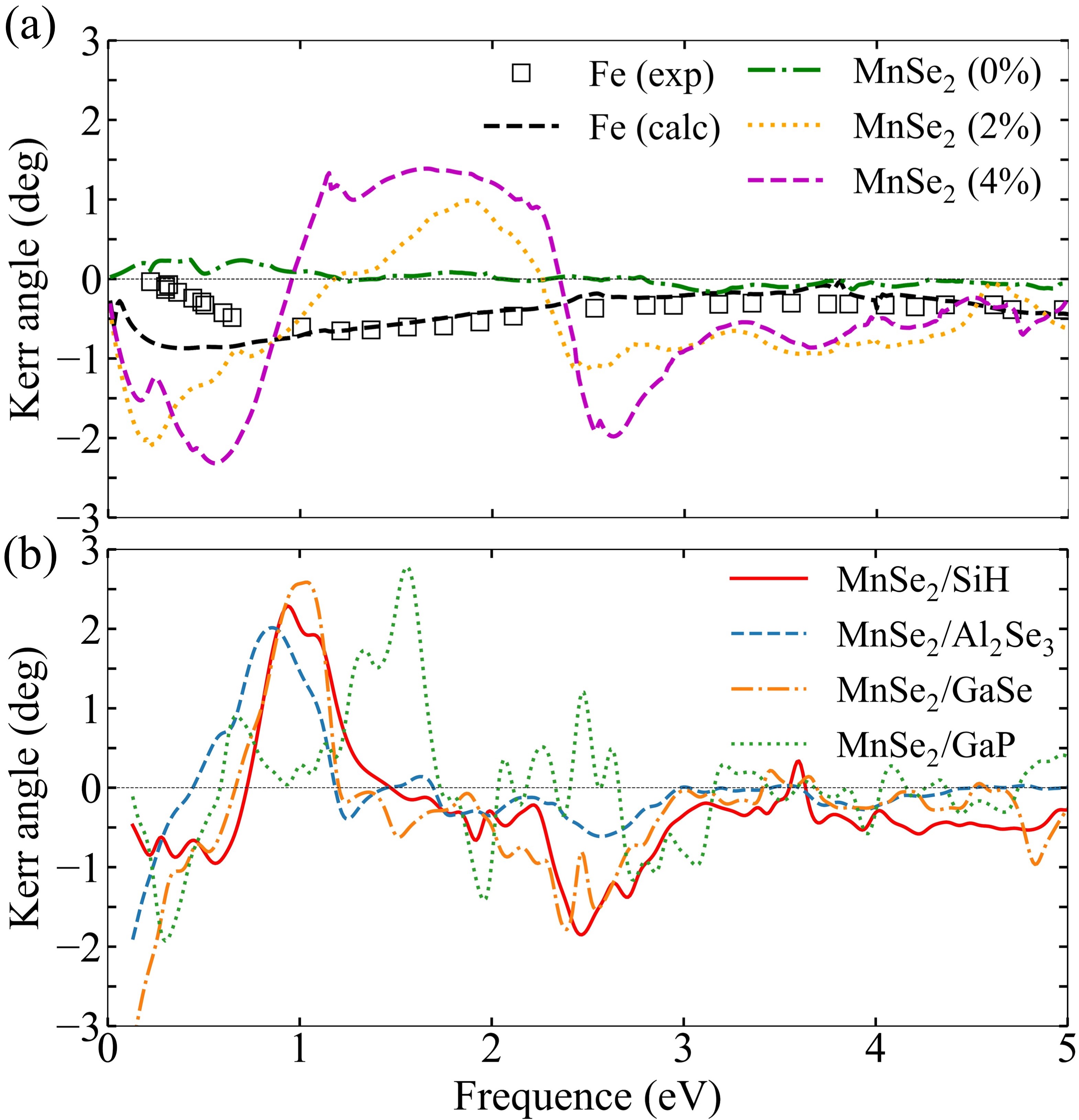}\\
		\caption{
			DFT results of Kerr rotation angle for MnSe$_2$ with 0\%, 2\%, 4\% tensile strain and heterostructures MnSe$_2$/X (X = SiH, Al$_2$Se$_3$, GaSe, and GaP).
			Experimental and numerical results of Fe \cite{Oppeneer1992} are also included for comparison.
		}\label{fig_kerr}
	\end{figure}
	
	We investigated the magneto-optical Kerr effect for MnSe$_2$ based structures.
	The Kerr rotation angle is given by:
	\begin{eqnarray}
		\begin{aligned}
			\theta_K(\omega)=Re\frac{\varepsilon_{xy}}{(1-\varepsilon_{xx})\sqrt{\varepsilon_{xx}}},
			\\
		\end{aligned}
		\label{eqs1}
	\end{eqnarray}
	where $\varepsilon_{xx}$ and $\varepsilon_{xy}$ are the diagonal and off-diagonal components of the dielectric tensor $\varepsilon$, and $\omega$ is the frequency of incident light.
	The dielectric tensor $\varepsilon$ can be obtained by the optical conductivity tensor $\sigma$ as $\varepsilon(\omega)=\frac{4{\pi}i}{\omega}\sigma(\omega)+I$, where I is the unit tensor.
	The calculated Kerr rotation angle as a function of photon energy for monolayer MnSe$_2$ with strain is shown in Fig. \ref{fig_kerr} (a), the result for heterostructures MnSe$_2$/X with X = GaP, GaSe, SiH and Al$_2$Se$_3$ is shown in Fig. \ref{fig_kerr} (b).
	According to our calculation results, there exist large Kerr rotation angles with out-of-plane magnetization, such as monolayer MnSe$_2$ with 2\% and 4\% tensile strain. 
	On the contrary, Kerr rotation angles with in-plane magnetization are small, such as monolayer MnSe$_2$ without strain.
	In addition, heterostructures MnSe$_2$/X with out-of-plane magnetization also have a large Kerr rotation angle.
	The experimental result for Fe \cite{Oppeneer1992} and our DFT result for Fe bulk are also included for comparison.
	The Kerr rotation angles for stretched MnSe$_2$ are about 4 times bigger than that of bcc Fe.
	Detailed results of Kerr rotation angles are given in Supplemental Material \cite{SM}.

    \section{Conclusion}
    Based on the DFT calculations, we studied the properties of monolayer MnSe$_2$ with strain and MnSe$_2$-based heterostructures.
    For monolayer MnSe$_2$, a metal-semiconductor transition happens at 2.2\% tensile strain, and an in-plane to out-of-plane MAE transition happens at 1.7\% tensile strain.
    In addition, $T_C$ of monolayer MnSe$_2$ increases with tensile strain.
    The heterostructures MnSe$_2$/X with X = GaP, GaSe, SiH, and Al$_2$Se$_3$ were studied, and the DFT calculation results show that they are room temperature ferromagnetic semiconductors with out-of-plane MAE.
    Large magneto-optical Kerr effect was found in monolayer MnSe$_2$ with a few percent tensile strain and MnSe$_2$-based heterostructures.
    Our results propose a way to obtain room temperature ferromagnetic semiconductors by metal-semicondutor transition in monolayer MnSe$_2$ by applying a few percent tensile stain or building heterostructures.

    \section {Acknowledgements}
    This work is supported by National Key R\&D Program of China (Grant No. 2022YFA1405100), National Natural Science Foundation of China (Grant No. 12074378), Chinese Academy of Sciences (Grants No. YSBR-030, No. JZHKYPT-2021-08, No. XDB33000000), Beijing Municipal Science and Technology Commission (Grant No. Z191100007219013).

\end{document}